\documentclass[prb,twocolumn,showpacs,preprintnumbers,amsmath,amssymb,superscriptaddress]{revtex4-2}

\usepackage{graphicx}
\usepackage{dcolumn}
\usepackage{bm}
\usepackage{color}
\usepackage{bbm}
\usepackage{subfigure}
\usepackage{natbib}

\usepackage[colorlinks=true,citecolor=blue,linkcolor=blue,urlcolor=blue]{hyperref}

\begin{document}

\title{Quantum phase transitions in a Dicke trimer with both photon and atom hoppings}

\author{Jun-Wen Luo}
\affiliation{School of Physics, Sun Yat-sen University, Guangzhou 510275, China}
\author{Bo Wang}
\email{bwang.physics@gmail.com}
\affiliation{Quantum Science Center of Guangdong-Hong Kong-Macau Great Bay Area, 3 Binlang Road, Shenzhen, China}
\affiliation{School of Physics, Sun Yat-sen University, Guangzhou 510275, China}
\author{Ze-Liang Xiang}
\email{xiangzliang@mail.sysu.edu.cn}
\affiliation{School of Physics, Sun Yat-sen University, Guangzhou 510275, China}
\date{\today}

\begin{abstract}
We investigate superradiant quantum phase transitions in a Dicke trimer model consisting of two types of hoppings, i.e., photon hoppings and atom hoppings. In the superradiant regime, the system can exist in two distinct phases: normal and frustrated superradiant phases, which are governed by both types of hoppings. Particularly, the interplay between these hoppings gives rise to interesting effects, such as triggering superradiance with much lower coupling strengths when both hoppings exhibit the same tendency. In contrast, with opposite tendencies, the competition between hoppings leads to a first-order phase transition between two different superradiant phases with translational symmetry broken. These findings enable the system to undergo a sequence of transitions across three phases by changing the coupling strength. Our work provides deep insights into competing interactions and quantum phase transitions in multi-cavity systems with geometric structures.
\end{abstract}
\maketitle

\section{Introduction}

Quantum phase transitions (QPTs) happen when a system undergoes an abrupt change of ground state due to the quantum fluctuation at zero temperature, leading to various exotic phenomena~\cite{sachdev1999quantum,RevModPhys.69.315,greentree2006quantum}. In recent years, QPTs have become an actively researched topic in quantum optics, arising intrinsically from light-matter interactions and revealing unique critical behaviors. As the effective description of light-matter coupling, the Dicke model consisting of a single mode cavity and an ensemble of two-level atoms can exhibit a superradiant QPT with the bosonic mode gaining a ground-state superradiance in the thermodynamic limit, i.e., the number of two-level atoms $N_a$$\rightarrow \infty$~\cite{PhysRevE.67.066203,HEPP1973360,PhysRevA.7.831,PhysRevA.75.013804,PhysRevLett.104.023601,PhysRevLett.104.130401,PhysRevLett.107.113602,PhysRevA.85.043821,PhysRevLettQED,PhysRevLett.131.113602}. Additionally, this QPT has been studied in the classical oscillator limit~\cite{RABIQPTN,RABI} in other cavity systems described by the quantum Rabi model~\cite{RABI,Rabi_Criti_PRL,JC} and optomechanical systems~\cite{PhysRevLettOPTO}.

Frustration, a fundamental phenomenon in condensed matter physics, prevents a system from reaching a state where all interactions are simultaneously satisfied~\cite{toulouse1987theory,moessner2006geometrical,PhysRevB.63.224401}. It typically arises from the competition between interactions within the system that favor different configurations, making it impossible to minimize the energy at the same time. A common example is the Heisenberg model~\cite{PhysRevLett.75.1823,PhysRevLett.121.107202,PhysRevB.41.9323}, where the competition between nearest- and next-nearest-neighbor antiferromagnetic couplings prevents neighboring spins from aligning antiparallel. Such an interaction competition causes the system to be highly degenerate or disordered, giving rise to exotic states of matter such as spin glasses and quantum spin liquids~\cite{Glass1,Glass2,balents2010spin,broholm2020quantum,RevModPhys.89.025003,RevModPhysIce,bramwell2001spinice,han2008geometric}. Recently, a unique superradiant QPT has been studied in the Dicke lattice with unusual critical behaviors due to the frustration~\cite{PhysRevLettFSPT,PhysRevResearch.5.L042016}. Coincidentally, analogous quantum critical phenomena have been studied in interconnected multi-cavity systems, such as the quantum Rabi triangle~\cite{PhysRevLettRTri,YYZ_TriCri} and Rabi ring~\cite{Ring0,PhysRevA.RING1,PhysRevARIng2} with similar geometric structures. In existing studies, anomalous quantum phases are induced by photon-driven correlations between cavities. However, an open question is whether introducing additional interactions can produce more complex correlations and uncover richer quantum phenomena.

\begin{figure}[tpb]
    \centering
    \includegraphics[width=0.7\linewidth]{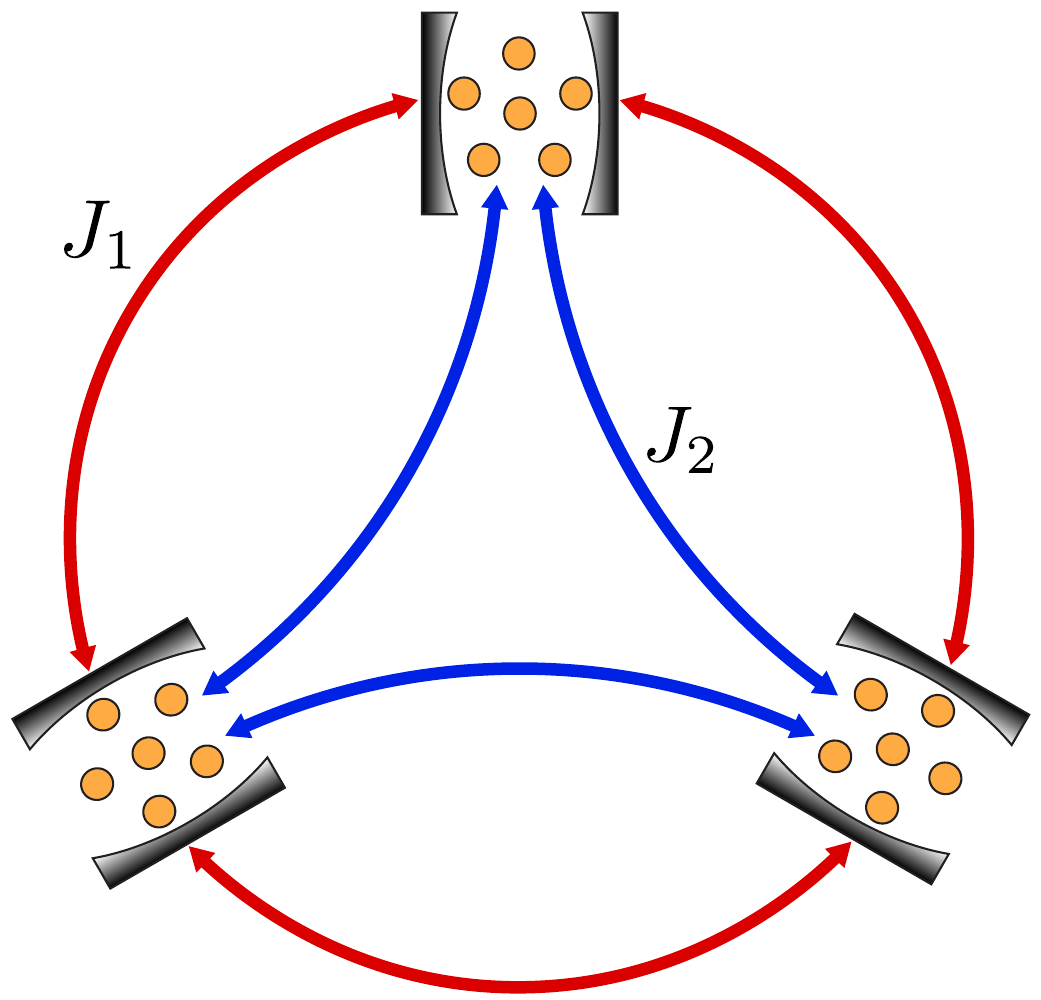}
    \caption{Schematic diagram of the Dicke trimer model with two types of hopping interactions. The red curve represents the photon hopping $J_1$ between neighboring Dicke sites, while the blue curve indicates the atom hopping $J_2$.}
    \label{fig:0}
\end{figure}

In this paper, we investigate a Dicke trimer model with two types of hoppings: photon hoppings $J_1$ and atom hoppings $J_2$, where each of two Dicke sites is connected through both hoppings, depicted in Fig.~\ref{fig:0}. Either type of hoppings impacts the determination of the ground state, and their interplay can be cooperative or competitive, depending on the sign and amplitude. Using the mean-field approach, we obtain the analytic solutions of the ground state. The system remains in the normal phase (NP) at weak coupling while it undergoes a superradiant QPT with $\mathbb{Z}_2$ symmetry broken when the coupling strength surpasses a critical threshold. In the superradiant regime, there exist two different phases: the normal superradiant phase (NSP) and the frustrated superradiant phase (FSP)~\cite{PhysRevLettFSPT}, which can be dominated by both hoppings. When the two hopping strengths have the same tendencies, their aligned interactions cause the system to go through a single superradiant phase. Therefore, the superradiant QPT can be achieved with reduced coupling strengths. However, when the hopping strengths have opposite tendencies, a first-order phase transition between the NSP and the FSP emerges in the superradiant regime. This leads to a three-phase transition (NP–NSP–FSP) driven merely by the coupling strength $g$ due to the competition between $J_1$ and $J_2$. We investigate the phase transitions for all values of ${J_1}$ and ${J_2}$, which can be rigorously categorized into six types, each represented as a distinct region on the $J_1$-$J_2$ plane. 

The paper is organized as follows. In Sec.~\ref{sec:model}, we study the Dicke trimer model with two hoppings and obtain the effective Hamiltonian. Sec.~\ref{sec:NormalPhase} diagonalizes the Hamiltonian and derives the critical points from the excitation spectrum. In Sec.~\ref{SuperradiantPhase}, the mean-field solutions for the ground state of superradiant phases are discussed. Then, in Sec.~\ref{Competition}, we show the phase diagram and discuss the interplay between ${J_1}$ and ${J_2}$. Finally, we give a summary in Sec.~\ref{Conclusion}.

\section{Model}
\label{sec:model}

We consider a Dicke trimer, where each site can be described as a single Dicke model, and neighboring sites are connected by both photon hopping interaction and atom hopping interaction. The total Hamiltonian of this system reads 
\begin{equation} \label{eq:1}
H=\sum_{i=1}^{3}H^{Dicke}_n+H_{hopping},
\end{equation}
where
\begin{equation} \label{eq:2}
H^{Dicke}_n=\omega a^\dag_n a_n+\Omega J^z_n+\frac{2\lambda}{\sqrt{N_a}}(a^\dag_n+a_n)J^x_n,
\end{equation}
\begin{equation} \label{eq:3}
H_{hopping}=\bar{J}_1 a^\dag_n a_{n+1} +\frac{\bar{J}_2}{N_a} J^+_n J^-_{n+1}+H.c,
\end{equation}
with a periodic boundary condition $a_{N+1}=a_1$. 

Here, $H^{Dicke}_n$ describes the behavior of the $n$th lattice site, where $a_n$ is the annihilation operator of the cavity with the frequency $\omega$, $N_a$ is the number of identical atoms in the ensemble with the frequency $\Omega$, described by collective spin operators $J^{x,z}_n$, and $\lambda$ denotes the local cavity-atom coupling strength. $H_{hopping}$ depicts the correlations between each two sites mediated by the photon hopping strength $\bar{J}_1$ and the atom hopping strength $\bar{J}_2$. The total Hamiltonian commutes with the parity operator $\Pi=exp[i\pi \sum_{n=1}^N(a^\dag_n a_n+J^z_n+\frac{N_a}{2})]$, which indicates a global $\mathbb{Z}_2$ symmetry. Furthermore, this total Hamiltonian also possesses translational symmetry, meaning that the $n$th lattice site doesn't depend on the site index $n$ in the symmetric phase. Considering the thermodynamic limit $N_a \rightarrow \infty$, the Dicke trimer model can undergo the superradiant phase transition with the strong atom-cavity interaction.

In order to separate the mean values and fluctuations under the thermodynamic limit, we apply a unitary transformation $U=\prod_{n=1}^N e^{-i\phi_n J^z_n }e^{-i\theta_n J^y_n}e^{\alpha_n\alpha^\dag_n-\alpha^*_n\alpha_n}$ to $H$. We obtain the effective Hamiltonian $\tilde{H}=U^\dag HU$ and then apply the Holstein-Primakoff transformation~\cite{HP} to the rotated collective spins operators. The transformations are given by $J^+_n=\sqrt{N_a-b^\dag_nb_n}b_n$, $J^-_n=b^\dag_n\sqrt{N_a-b^\dag_nb_n}$ and $J^z_n=(N_a/2)-b^\dag_nb_n$, where $[b_n,b_n^\dag]=1$, mapping the collective spin operators to bosonic operators. The transformed Hamiltonian is obtained 
\begin{equation} \label{eq:4}
\tilde{H}=H_{q}+H_{l}+E_{GS},
\end{equation}
where $H_q$ is the quadratic Hamiltonian, $H_l$ is the linear Hamiltonian, and $E_{GS}$ is the ground-state energy. For simplicity, we rescale the energy by $g=2\lambda/\sqrt{\omega\Omega}$, $\bar{\alpha}_n=\sqrt{\omega/N_a\Omega}$ and obtain
\begin{equation} \label{eq:6}
\begin{split}
 \bar{E}_{GS} =&\,\sum^N_{n=1}(\bar{\alpha}_n^2+\frac{1}{2}\cos\theta_n+g\bar{\alpha}_n\sin\theta_n\cos\phi_n\\
&\,+2J_1\bar{\alpha}_n\bar{\alpha}_{n+1}+
\frac{1}{2}J_2\sin\theta_n\sin\theta_{n+1}\cos\phi_n\cos\phi_{n+1}),
\end{split}
\end{equation}
where $\bar{E}_{GS}=E_{GS}/N_a\Omega$, $J_1=\bar{J}_1/\omega$ and $J_2=\bar{J}_2/\Omega$. Globally minimizing $\bar{E}_{GS}$ with respect to $\bar{\alpha}_n$ and $\theta_n$,
\begin{equation}\label{eq:7_7}
\frac{\partial\bar{E}_{GS}}{\partial\bar{\alpha}_n}=2\bar{\alpha}_n+g\sin\theta_n\cos\phi_n+2J_1(\bar{\alpha}_{n-1}+\bar{\alpha}_{n+1})=0,
\end{equation}
\begin{equation} \label{eq:8}
\begin{split}
\frac{\partial\bar{E}_{GS}}{\partial\theta_n} =&\,-\frac{1}{2}\sin\theta_n+g\bar{\alpha}_n\cos\theta_n\cos\phi_n+\frac{1}{2}J_2\cos\theta_n\cos\phi_n\\
&\,\times(\sin\theta_{n-1}\cos\phi_{n-1}+\sin\theta_{n+1}\cos\phi_{n+1})=0,
\end{split}
\end{equation}
now the mean-field ground state energy $\bar{E}_{GS}(\{\bar{\alpha}_{n}\})$ can be found with cavity degrees of freedom only. For ease of analysis, we set a new type of parameter $x_n$ to replace coherence $\bar{\alpha}_n$, namely, 
\begin{equation} \label{eq:9}
x_n=\bar{\alpha}_n+J_1(\bar{\alpha}_{n-1}+\bar{\alpha}_{n+1})
\end{equation}
which is a linear combination of $\bar{\alpha}$.
Finally, the ground state energy can be given as 
\begin{equation} \label{eq:10}
\bar{E}_{GS}(\{x_{n}\}) =\sum_{i=1}^{3} \tilde{C}x_n^2-\frac{1}{2}\sqrt{1-\frac{4x_n^2}{g^2}}
+2\tilde{B} x_n x_{n+1},
\end{equation}
with $\tilde{C}=\frac{1+J_1}{(-1+J_1)(1+2J_1)}$ and  $\tilde{B}=(\frac{J_1}{1+J_1-2J_1^2}+\frac{J_2}{g^2})$.

\section{Normal Phase}\label{sec:NormalPhase}

The global minimum of ground state energy $\bar{E}_{GS}$ gives rise to the solution $\{\alpha_n,\theta_n,\phi_n\}$, which causes $H_l=0$. We perform a Fourier transform, i.e., $a^\dag_n=\sum_ke^{ikn}a^\dag_k/\sqrt{N}$ and $b^\dag_n=\sum_ke^{ikn}b^\dag_k/\sqrt{N}$, to the Hamiltonian with $k=0,\pm2\pi/3.$ As the cavity coherence $\alpha_n=0$, we have the effective Hamiltonian for the normal phase, and it reads
\begin{equation} \label{eq:11}
H_{np}=\sum_k\omega_k a^\dag_ka_k+\Omega_k b^\dag_kb_k-\lambda(a^\dag_{-k}+a_k)(b^\dag_{-k}+b_k)
\end{equation}
where $\omega_k=\omega+2\bar{J}_1\cos k$, $\Omega_k=\Omega+2\bar{J}_2\cos k$. After diagonalization, the excitation energies are $\varepsilon^{(\pm)}_k$ (see in Fig.~\ref{WTF}). The critical point of the system can be derived from the excitation energy $\varepsilon^{-}_k$ vanishing at
\begin{equation} \label{eq:12}
g_c=min\{\sqrt{(1 + 2 J_1 \cos k)(1 + 2 J_2\cos k)}|k\},
\end{equation}
with $k=0,\pm2\pi/3$, which indicates that there are two kinds of critical points, including zero-momentum $k=0$ and finite-momentum $k=\pm2\pi/3$ modes. It is worth noting that the critical coupling strength $g_c$ is obviously dominated by two hopping strength $\bar{J}_{1/2}$, so different values of $J_1$ and $J_2$ will engender different critical points. In $J_1$-$J_2$ space (Fig.~\hyperref[fig:1]{5(b)}), a curve,
\begin{equation} \label{eq:12-1}
J_1=-J_2/(1+J_2),
\end{equation}
has been provided to partition several regions, where the system will experience distinct phase transitions. The area above the curve corresponds to $g_c^{+}=\sqrt{(1-J_1)(1-J_2)}$ with finite-momentum $k=\pm2\pi/3$ as well as the area below it corresponds to $g_c^{-}=\sqrt{(1+2J_1)(1+2J_2)}$ with the zero-momentum $k=0$. Here, we consider only the small hopping rates $-1/2<J_{1/2}<1/2$. 

\section{Superradaint Phase}\label{SuperradiantPhase}

\begin{figure}
    \centering    \includegraphics[width=0.49\textwidth]{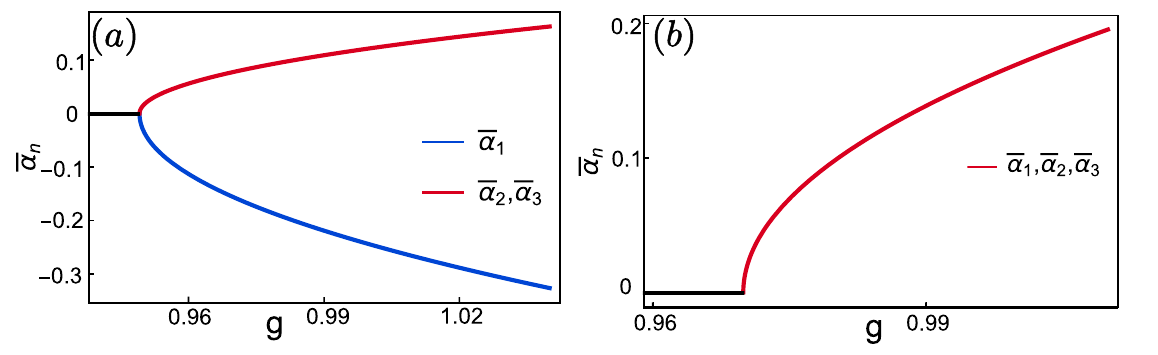}
    \caption{(a) One of the six degenerate ground-state solutions for the renormalized coherence of the cavity in the $n$th cavity. The red solid line corresponds to two sites with the same coherence as a pair, and the blue solid line corresponds to another with different sign and amplitude. (b) One of the two degenerate ground-state solutions. The red solid line represents the non-zero superradiance, with all sites having identical coherence.}
    \phantomsection
    \label{fig:2a}
    \phantomsection
    \label{fig:2b}
\end{figure}

\subsection{Superradint Frustrated Solution}

In the case where both the matter and photon hopping strength are positive, the Dicke trimer will undergo a frustrated superradiant phase transition~\cite{PhysRevLettFSPT}, driven by the combined effect of both factors acting in the same direction. The ground state of the frustrated superradiant phase prefers a coherence $\alpha$ with opposite signs for neighboring sites, resulting in a six-fold degeneracy. This mechanism is similar to the antiferromagnetic phase in the spin-$\frac{1}{2}$ triangular Ising model, where the coherence $\alpha$ can be mapped to the spin. However, unlike spins that undergo only sign changes while the angular momentum remains fixed, $\alpha$ can vary in both sign and amplitude, leading to different macroscopic occupations at each Dicke site. 

\begin{figure}
    \centering    \includegraphics[width=0.49\textwidth]{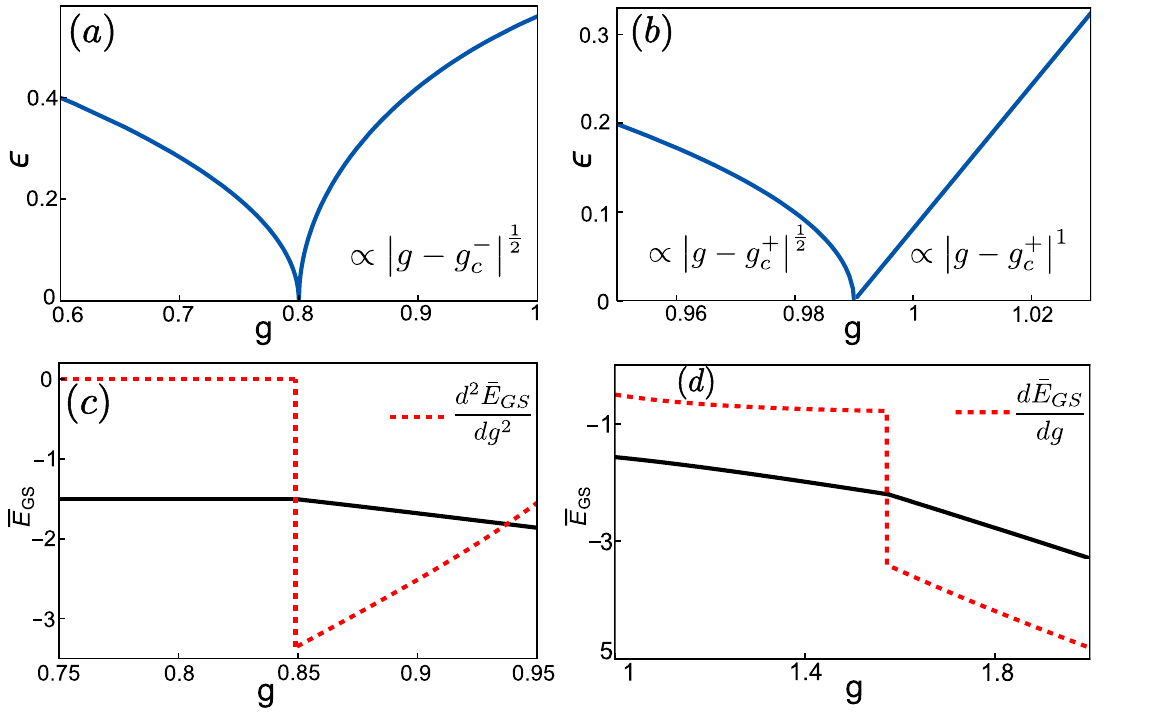}
    \caption{Top panel: excitation energies as a function of the coupling strength $g$ for (a) $g_c^-$ and (b) $g_c^+$. In (b), the critical modes have distinct power laws with exponents $\gamma=1$, which is a noticeable indication of the frustrated superradiant phase. Bottom panel: (c) the rescaled ground-state energy and its second-order derivative (red dashed line) as a function of the coupling strength $g$ for indicating the ground-state energy has a discontinuity in its second-order derivative with respect to $g$; (d) the rescaled ground-state energy and its first-order derivative as a function of the coupling strength $g$. The discontinuity signifies a first-order phase transition at $g_L$.}
    \label{WTF}
\end{figure}

Using the global minimum condition for the ground-state energy, we transform the equation $\partial\bar{E}_{GS}/\partial\theta_n=0$ with the solution of $\partial\bar{E}_{GS}/{{\partial\bar{\alpha}_n}}=0$ and apply the parameter $x$ to replace $\alpha$ for conciseness. By employing the monotonic method (Appendix.~\hyperref[Appendix B]{B}), we find out the configuration of $x$ is that two sites have the same sign and the remaining site has the opposite sign, i.e, $x_n<0,x_{n-1}=x_{n+1}>0$, featuring sixfold degeneracy. This configuration is valid only if the coefficient $\tilde{B}$ in Eq.~\ref{eq:10} is positive corresponding to the critical coupling strength $g_c^+$. Using this configuration, we can reduce Eq.~\eqref{eq:8} to
\begin{equation} \label{eq:14}
\frac{2J_2x_1}{g^2}+\frac{x_2}{g^2\sqrt{1-\frac{4x_2^2}{g^2}}}+\frac{x_2+J_1(-2x_1+x_2)}{(-1+J_1)(1+2J_1)}=0,
\end{equation}
\begin{equation} \label{eq:15}
\frac{x_1-J_1x_2}{1+J_1-2J_1^2}-\frac{x_1}{g^2\sqrt{1-\frac{4x_1^2}{g^2}}}-\frac{J_2(x_1+x_2)}{g^2}=0.
\end{equation}
Since the equations above cannot be analytically solved, we find an asymptotic solution near the critical point $g_c^+$:
\begin{equation} \label{eq:16}
x_1\simeq -\frac{2\sqrt{(1-J_2)g_c^+}}{\sqrt{3}}|g-g_c^+|^{\frac{1}{2}}+O((g-g_c^+)^{\frac{3}{2}}),
\end{equation}
\begin{equation} \label{eq:17}
x_2\simeq x_3\simeq \frac{\sqrt{(1-J_2)g_c^+}}{\sqrt{3}}|g-g_c^+|^{\frac{1}{2}}+O((g-g_c^+)^{\frac{3}{2}}).
\end{equation}
By substituting back with the linear relation Eq.~\eqref{eq:9}, we obtain the mean values of cavity coherence,
\begin{equation} \label{eq:18}
\bar{\alpha}_1\simeq \frac{2\sqrt{(1-J_2)g_c^+}}{\sqrt{3}(-1+J_1)}|g-g_c^+|^{\frac{1}{2}}+O((g-g_c^+)^{\frac{3}{2}}),
\end{equation}
\begin{equation} \label{eq:19}
\bar{\alpha}_2= \bar{\alpha}_3\simeq  \frac{\sqrt{(1-J_2)g_c^+}}{\sqrt{3}(1-J_1)}|g-g_c^+|^{\frac{1}{2}}+O((g-g_c^+)^{\frac{3}{2}}),
\end{equation}
which is plotted in Fig.~\hyperref[fig:2a]{2(a)}. Based on the configuration of the coherence $\alpha$ at each site, the excitation energies $\epsilon(g)$ can be obtained as a function of the coupling strength $g$, as shown in Fig.~\hyperref[WTF]{3(b)}, where the critical exponent changes from $\mu=1/2$ to $\mu=1$ around $g_c^+$ indicating the frustrated superradiant phase transition. 

\subsection{Non-Frustrated Superradiant Solution}

When both matter and photon hopping strength are negative, the Dicke trimer undergoes a transition into a normal superradiant phase. Each site now exhibits the same coherence $\alpha$ resulting in completely identical occupations across all sites. It is mapping to the ferromagnetic phase in the spin-$\frac{1}{2}$ triangular Ising model, where the spins on all sites are identical with translational symmetry.

By applying the Cauchy-Schwarz inequalities (see Appendix.~\hyperref[Appendix C]{C}), the ground state corresponding to the global minimum of system energy can be determined by minimizing $\bar{E}_{GS}$ with $\bar{\alpha}_1=\bar{\alpha}_2=\bar{\alpha}_3$. The coherence for each site is
\begin{equation} \label{eq:23}
 \bar{\alpha}_{1,2,3}=\pm\frac{1}{2} g\sqrt{\frac{1}{(1 + 2 J_1)^2} -\frac{1}{(g^2 - 2 (J_2 + 2 J_1 J_2))^2}},
\end{equation}
with two-fold degeneracy, shown in Fig.~\hyperref[fig:2a]{2(b)}. The normal superradiant phase transition of the Dicke trimer exhibits an identical nature to that of the superradiant phase transition in a single Dicke model. Its excitation energy function $\epsilon(g)$, depicted in Fig.~\hyperref[WTF]{3(a)}, has critical exponent  $\mu=\frac{1}{2}$, which is a commonplace characteristic for the superradiant phase transition.

\begin{figure}
    \centering
    \includegraphics[width=1\linewidth]{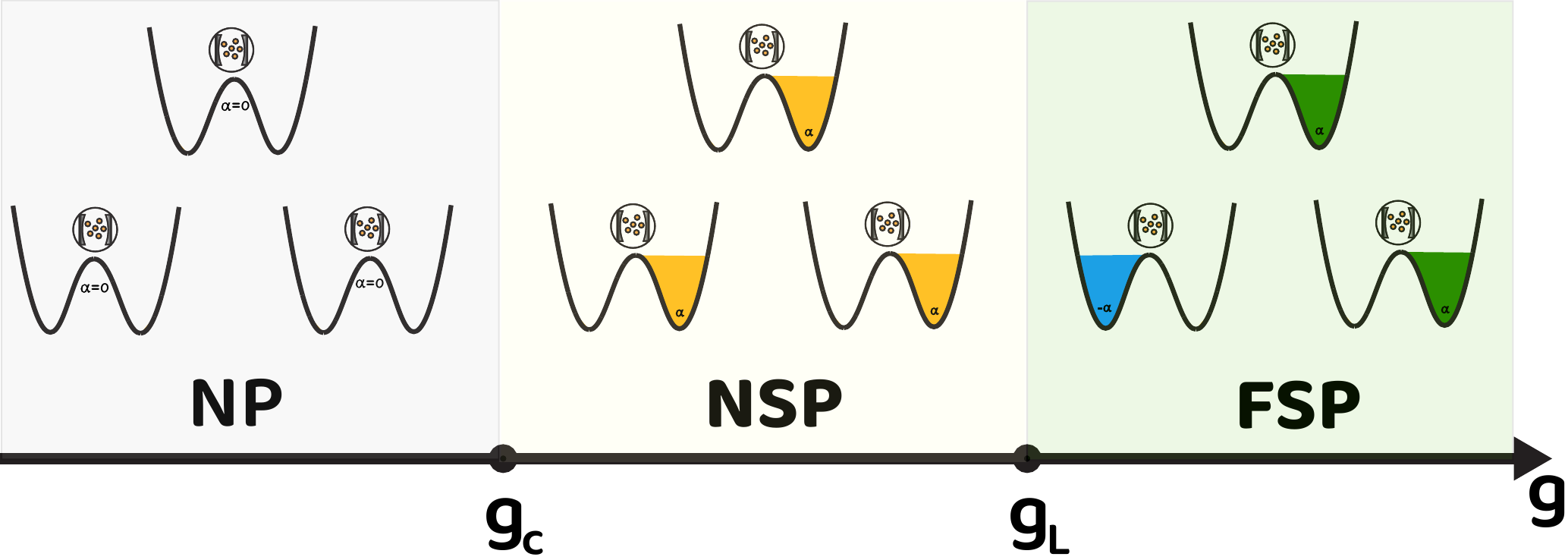}
    \caption{A sequence of transitions across the three phases of the Dicke trimer model. The system will experience a second-order phase transition with $\mathbb{Z}_2$ symmetry broken at $g_c$ and then undergo a first-order phase transition with translational symmetry broken around $g_L$ with the coupling strength $g$ increasing.}
    \label{fig:22}
\end{figure}

\section{Cooperation and Competition Between two hoppings}\label{Competition}

In the Dicke trimer model with only photon hopping~\cite{PhysRevLettFSPT}, which can be described by the Hamiltonian Eq.~\eqref{eq:1} with $J_2=0$, the photon hopping makes neighboring cavities correlated. In this case, the mean-field theory shows that the sign of $J_1$ significantly affects the minimization of the system ground-state energy and dominates the quantum phases: $J_1<0$ induces the NSP, while $J_1>0$ leads to the FSP. Thus, for a given sign of $J_1$, increasing the coupling strength $g$ can drive the system from the normal phase to a nontrivial quantum phase (normal or frustrated superradiant phase).

When atom hoppings are introduced into the system, it becomes more complicated and interesting. This is because the system's correlated behaviors will arise not only in neighboring cavities but also in neighboring atom ensembles. Considering the strong interaction between the cavity mode and atoms, it is easy to understand that these correlations will manifest as deeply intertwined characteristics within the Dicke trimer system, leading to even more fascinating critical behaviors.

\subsection{First-order phase transition with translational Symmetry Broken}

For the Hamiltonian~\eqref{eq:1}, we find that the sign of 
\begin{equation} \label{eq:24-1}
\tilde{B}=(\frac{J_1}{1+J_1-2J_1^2}+\frac{J_2}{g^2})
\end{equation}
in the ground-state energy of Eq.~\eqref{eq:10} play a key role in determining the mean-field value $\alpha_n$ of neighboring sites. When $\tilde{B}>0$, opposite signs (antialigned) for $\alpha_n$ at neighboring sites are favored to achieve the global minimum of the ground-state energy. However, for the lattice geometry shown in Fig.~\ref{fig:0}, it is impossible to satisfy the antialigned configuration, meaning the system will tend to a frustrated state. When $\tilde{B}<0$, all sites are allowed to be aligned to realize the global minimum, so the system will enter into the superradiant phase without frustration.

Obviously, the point $\tilde{B}=0$ marks the boundary between the frustrated phase and the normal superradiant phase, associated with a first-order phase transition. In a single-hopping model~\cite{PhysRevLettFSPT,PhysRevResearch.5.L042016}, this phase transition is only induced by a hopping parameter $J$. Conversely, in the two-hopping model, the equation~\eqref{eq:24-1} indicates that such a phase transition can be triggered not only by the hopping parameter $J$ but also by the coupling strength $g$. Importantly, the latter implies that the system can undergo two distinct types of phase transitions as $g$ increases: a second-order phase transition at $g_c$ and a first-order phase transition point at
\begin{equation} \label{eq:25}
g_L=\sqrt{\frac{(-1-J_1+2J_1^2)J_2}{J_1}},
\end{equation}
as shown in Fig.~\ref{fig:22}.  


\subsection{Anomalous Phase Diagram}

Based on the above results, an analytical phase diagram of the Dicke trimer is presented in Fig.~\ref{fig:1}. We first show the phase diagram in the $g-J_2$ parameter space (see in Fig.~\hyperref[fig:1]{5(a)}) with the photon hopping $J_1$ fixed. As depicted, the system can undergo a phase transition from the NP to the NSP (FSP) through second-order critical lines, as well as a transition from NSP to FSP via a first-order line. There is a triple point at the intersection of these critical lines, where three phases coexist. On the left-hand side of the triple point, the system undergoes a three-phase (NP-NSP-FSP) transition as the coupling strength $g$ increases, which is consistent with our deduction.

We investigate the phase transitions driven by the coupling strength $g$ across all values of $J_1$ and $J_2$. These transitions can be rigorously classified into six types, corresponding to six distinct regions in the $J_1$-$J_2$ plane, as shown in Fig.~\hyperref[fig:1]{5(b)}. The six regions are separated by the coordinate axes and the dividing line $J_{1}=-J_{2}/(1+J_{2})$.

In fact, this anomalous phase diagram results from the joint effect of the photon hopping $J_1$ and the atom hopping $J_2$. In the following, we will discuss the critical behaviors generated by it in detail.


\subsection{Interplay between $J_1$ and $J_2$}

\begin{figure}
    \centering
    \includegraphics[width=0.8\linewidth]{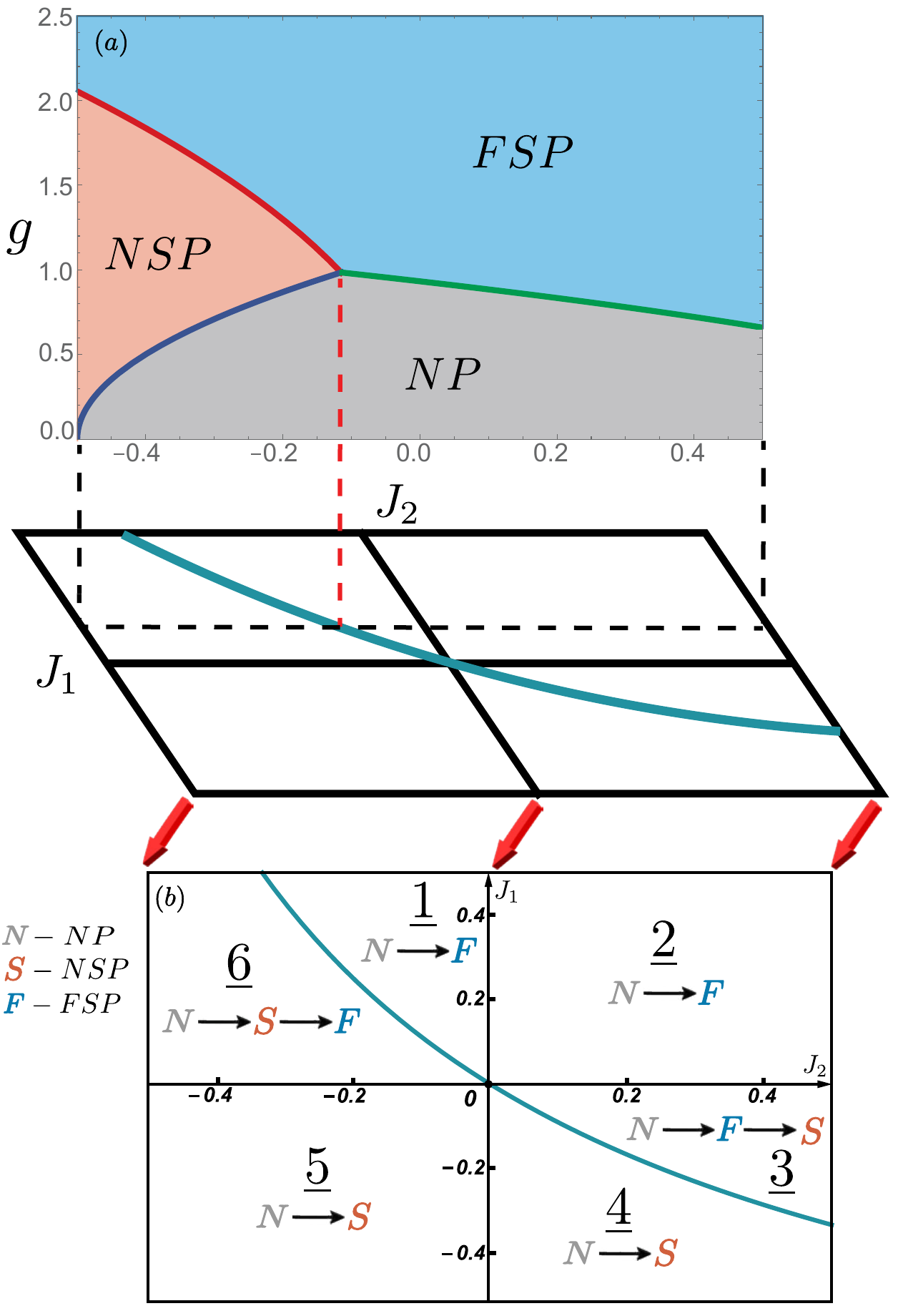}
    \caption{(a) The phase diagram in the $g-J_2$ parameter space.  
    The solid green (blue) line represents the second-order phase boundary associated with the critical point $g^+_c$ ($g^-_c$), and the solid red line stands for the first-order phase boundary related to $g_L$. We set $J_1=0.1$. The phase transitions induced by $g$ for all values of $J_1$ and $J_2$ is depicted in the $J_1-J_2$ plane (b), classified as six distinct regions labeled from 1 to 6. The green curve $J_{1}=-J_{2}/(1+J_{2})$ is the dividing line where $g_c^+$=$g_c^-$. The area above the curve corresponds to $g_c^{+}$, and the area below it corresponds to $g_c^{-}$. The six regions are divided by the axis-$J_1$, the axis-$J_2$, and the divided curve line. }
    \label{fig:1}
\end{figure}

\begin{table*} 
\centering
\caption{Classifications of phase transitions induced by the coupling $g$ in the $J_1$-$J_2$ parameter space.}
\begin{tabular}{|c|c|c|c|c|c|c|}
\hline
Region & 1 & 2 & 3 & 4 & 5 & 6 \\
\hline
Hoppings & $J_1>0$, $J_2<0$ & $J_1>0$, $J_2>0$ & $J_1<0$, $J_2>0$ & $J_1<0$, $J_2>0$ & $J_1<0$, $J_2<0$ & $J_1>0$, $J_2<0$ \\
\hline
Phase transition &$NP \rightarrow FSP$ &$NP \rightarrow FSP$ &$NP \rightarrow FSP \rightarrow NSP$ & $NP\rightarrow NSP$ & $NP\rightarrow NSP$ & $NP \rightarrow NSP \rightarrow FSP$ \\
\hline
$g_L$ & $g_L<g_c^+$ & non-existent & $g_c^+<g_L$ & $g_L<g_c^-$ &  non-existent & $g_c^-<g_L$ \\
\hline
\end{tabular}
\end{table*}

 Within the single-hopping framework, the sign of the hopping strength $J_{1/2}$ will determine the system's quantum phase to be FSP ($J_{1/2}>0$) or NSP ($J_{1/2}<0$). When both photon hopping and atom hopping are included, there is a cooperative or competitive effect on the configuration of the ground state of the Dicke trimer, as implied by Eq.~\eqref{eq:24-1}. In particular, when the signs of two hopping strengths are different, there will be two completely different trends (one favors FSP, while the other favors NSP) competing with each other, making the critical behavior of the system anomalous.

We first discuss the $J_1$-$J_2$ cooperative effect on the Dicke trimer model, as shown by regions 2 and 5 in the $J_1$-$J_2$ plane (see in Fig.~\hyperref[fig:1]{5(b)}). In these regions, the value of the hopping strengths $J_1$ and $J_2$ have the same sign, indicating that the photon hopping and the atom hopping contribute the same tendency to the formation of quantum phases. For the example of region 2 ($J_1>0$ and $J_2>0$), $\tilde{B}$ in Eq.~\eqref{eq:24-1} is larger than 0, indicating that the system will undergo the phase transition between the NP and the FSP by increasing the coupling strength $g$. More importantly, from the corresponding critical point $g_c^{+}=\sqrt{(1-J_1)(1-J_2)}$ with $0<J_{1/2}<1/2$, we find that the required coupling strength to reach the phase transition can be significantly reduced. The same mechanism also applies to $J_1<0$ and $J_2<0$ in region 5. 

When the signs of two hopping strengths are opposite, the $J_1$-$J_2$ competitive effect becomes the dominant theme, where the trend of breaking translational symmetry ($J_{1/2}>0$)  competes with the trend of preserving it ($J_{2/1}<0$). The former induces the FSP, while the latter leads to the NSP. From the previous discussion, we have identified a clear boundary, $J_{1}=-J_{2}/(1+J_{2})$ that completely divides the region of broken translational symmetry from that of preserved translational symmetry, as illustrated by the curve in Fig.~\hyperref[fig:1]{5(b)}. In region 1, broken translational symmetry dominates the ground state energy, resulting in a phase transition from the NP to the FSP as the coupling strength $g$ increases. In contrast, region 4 tends to preserve translational symmetry, allowing the NSP to be induced.

Interestingly, as the coupling strength $g$ continues to increase, the original result of the competition can be reversed, leading to the emergence of an additional phase transition point $g_L$ alongside the critical point $g_c$, as shown in Fig.~\ref{fig:22}. These phenomena occur in regions 3 and 6 in Fig.~\ref{fig:1}. In the example of region 6, the system undergoes a second-order phase transition from the NP to the NSP and then enters into the FSP with a first-order phase transition. This is intriguing because the light-matter interaction term typically governs the characteristics of the superradiant phase with $\mathbb{Z}_2$ symmetry broken, while the hopping term induces the broken translational symmetry. The latter determines whether the critical phenomena of the former are exotic (frustrated) or not (non-frustrated). In a two-hopping model, this scenario has been changed because the coupling strength can directly trigger the phase transition between the FSP and the NSP, as indicated by $g_L$~\eqref{eq:25}. To our knowledge, this is the first time to show that there are two transition points featured by the coupling strength ($g_c$ and $g_L$) in the study of superradiant phase transitions, raising the question of why the coupling parameter $g$ governing the light-matter interaction term can induce both $\mathbb{Z}_2$ symmetry breaking and translational symmetry breaking.

It is worth noting that this situation occurs only when $g_c<g_L$, because the critical point $g_c$ determines whether the system can enter into a quantum phase with a macroscopic occupation. As a result, the boundary line, $J_{1}=-J_{2}/(1+J_{2})$, separating the regions of broken translational symmetry and preserved translational symmetry, shifts to the horizontal line $J_{1}=0$, as illustrated in Fig.~\hyperref[fig:1]{5(b)}. 


\section{Summary}\label{Conclusion}

We investigate quantum phase transitions in a Dicke trimer model with both photon and atom hoppings connecting each single site. These two types of hoppings have a combined effect on determining the ground state and compete against each other when they exhibit different tendencies. We obtain the analytical solution for the ground state of the system and identify the quantum phase transition within it.

When the photon hopping $J_1$ and atom hopping $J_2$ share the same sign, indicating a similar tendency in determining the system's ground state, the system only undergoes a second-order phase transition from normal phases to superradiant phases with $\mathbb{Z}_2$ symmetry broken. The system's ground state undergoes a phase transition from the NP to the NSP with negative hopping strength and from the NP to the FSP with positive one. When the photon hopping $J_1$ and atom hopping $J_2$ have the opposite tendency, their competition lead to a first-order phase transition between the NSP and the FSP with translational symmetry broken. In this scenario, the system can undergo sequential phase transitions and exhibit three distinct phases by just tuning the coupling strength $g$. This also indicates coupling strength g can not only trigger the broken $Z_2$ symmetry but also the broken translational one. We obtain the phase diagram and find out the phase transitions for all possible values of $J_1$ and $J_2$ can be classified into six types, corresponding to six different regions in the $J_1$-$J_2$ space.

Our work provides deep insight into the correlations among interconnected multi-cavity systems in a complex situation and enhances the understanding of QPTs in light-matter coupling systems with geometric structures. This framework also presents a fresh perspective to investigate the competing interactions within the lattice system. 

\section*{Acknowledgments}

This work is supported by the National Natural Science Foundation of China (Grant No. 12375025 and No. 11874432).


\appendix
\section{Diagonalization}

We apply the unitary transformation $U=\prod_{n=1}^N e^{-i\phi_n J^z_n }e^{-i\theta_n J^y_n}e^{\alpha_n\alpha^\dag_n-\alpha^*_n\alpha_n}$ to the original Hamiltonian and obtain the effective Hamiltonian with the explicit form, 

\begin{equation} \label{eq:A1}
\begin{split}
\tilde{H}& =\sum_{i=1}^{3}\{\omega (a^\dag_n+\alpha_n^*)(a_n+\alpha_n)+\Omega(-\sin\theta_n J^x_n+\cos\theta_n J^z_n)
\\&+\frac{2\lambda}{\sqrt{N_a}}(a_n+a^\dag_n+2\alpha_n)(\cos\theta \cos\phi_n J^x_n-\sin\phi_n J^y_n
\\&+\sin\theta_n \cos\phi_n J_n^z)+\bar{J}_1[(a^\dag_n+\alpha_n^*)(a_n+1+\alpha_n+1)
\\&+ (a_n+\alpha_n)(a^\dag_{n+1}+\alpha_{n+1}^*)]
+\frac{2\bar{J}_2}{N_a}[(\cos\theta_n \cos\phi_n J_n^x\\&
-\sin\phi_n J^y_n+\sin\theta_n \cos\phi_n J^z_n)(\cos\theta_{n+1} \cos\phi_{n+1} J_{n+1}^x\\&
+\cos\phi_{n+1} \sin\theta_{n+1} J_{n+1}^z-\sin\phi_{n+1} J_{n+1}^y)+
(\cos\phi_n J^y_n
\\&+\sin\phi_n \cos\theta_n J^x_n+\sin\phi_n \sin\theta_n J^z_n)(\cos\phi_{n+1}J_{n+1}^y
\\&+\cos\theta_{n+1} \sin\phi_{n+1 }J_{n+1}^x+\sin\theta_{n+1} \sin\phi_{n+1} J_{n+1}^z)]\}.
\end{split}
\end{equation}
Then, we apply the HP transformation to the collective spin operators and we have $J^+_n=\sqrt{N_a}b_n$, $J^-_n=\sqrt{N_a}b^\dag_n$, and $J^z_n=(N_a/2)-b^\dag_nb_n$ in the thermodynamic limit $N_a \rightarrow \infty$. The effective Hamiltonian reads as, which can be separated into three parts, the linear term $H_l$, the quadratic term $H_q$, and the ground state energy $E_{GS}$. The ground state energy $E_{GS}$ reads as
\begin{equation} \label{eq:A3}
\begin{split}
E_{GS}& =\sum_{i=1}^{3}[\omega \alpha^2_n+\frac{N_a}{2}\Omega \cos\theta_n+2\sqrt{N_a}\lambda\alpha_n \sin\theta_n \cos\phi_n
\\&+2\bar{J}_1\alpha_n\alpha_{n+1}+\frac{\bar{J}_2}{2N_a}\sin\theta_n \sin\theta_{n+1} \cos\phi_n \cos\phi_{n+1}],
\end{split}
\end{equation}
with the $Im(\alpha_n)=0$ and $\cos\theta_n$ being negative for the the minimum of energy. The linear term, 
\begin{equation} \label{eq:A3}
\begin{split}
H_l& =\sum_{i=1}^{3}\left[A_n(a_n+a^\dagger_n)+B_n(b_n+b^\dagger_n)\right],
\end{split}
\end{equation}
vanishes where $A_n=\omega\alpha_n-2\sqrt{N_a}\sin\theta_n \cos\phi_n+\bar{J}_1(\alpha_{n-1}+\alpha_{n+1})$ and $B_n=-\frac{1}{2}\sqrt{N_a}\Omega \sin\theta_n-2\lambda \alpha_n \cos\theta_n+\frac{1}{2}\bar{J}_2\sqrt{N_a}\cos\phi_n \cos\phi_{n+1}\sin(\theta_n+\theta_{n+1})$ due to the global minimum of $E_{\rm GS}$. By using the quadratures $q_n=(a_n+a^\dagger_n)/\sqrt{2}$, $p_n=i(a_n-a^\dagger_n)/\sqrt{2}$, $Q_n=(b_n+b^\dagger_n)/\sqrt{2}$, and $Q_n=i(b_n-b^\dagger_n)/\sqrt{2}$, the quadratic term is given by
\begin{equation} \label{eq:A4}
\begin{split}
H_q& =\sum_{i=1}^{3}\left[\frac{\omega}{2}(q_n^2+p_n^2)-\frac{\Omega}{2\cos\theta_n}(Q_n^2+P_n^2)+2\lambda \cos\theta_n\right.\\& cos\phi_n Q_n q_n+\bar{J}_1(q_n q_{n+1}+p_n p_{n+1})+\bar{J}_2(P_n P_{n+1}\\& \left.+\cos\theta_n \cos\theta_{n+1}Q_n Q_{n+1})\right].
\end{split}
\end{equation}

According to Willamson's theorem~\cite{serafini2023quantum}, one can use the symplectic transformation to decompose the quadratic Hamiltonian $H_q$ and obtain its eigenvalues of the Hamiltonian matrix. The first-excitation energy around critical points is shown in Fig.~\ref{WTF}.


\section{Monotonic Method}\label{Appendix B}

\begin{figure}
    \centering
    \subfigure[]{
        \includegraphics[width=0.225\textwidth]{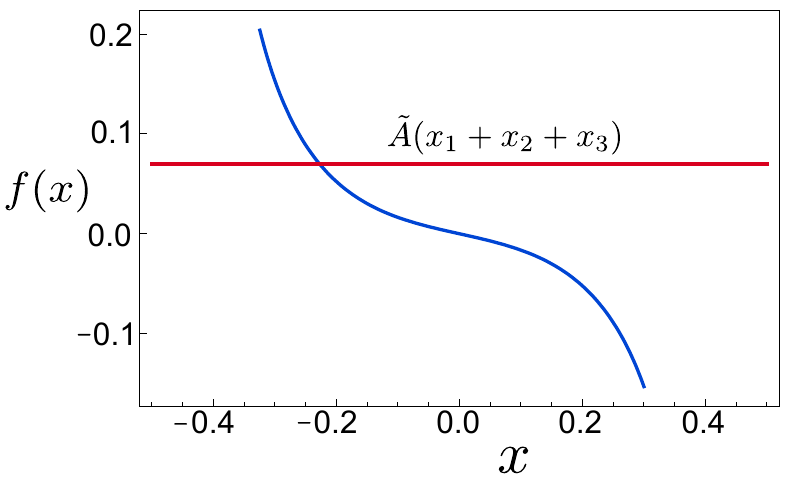}
        \label{fig:figure2}
    }
    \hfill
    \subfigure[]{
        \includegraphics[width=0.225\textwidth]{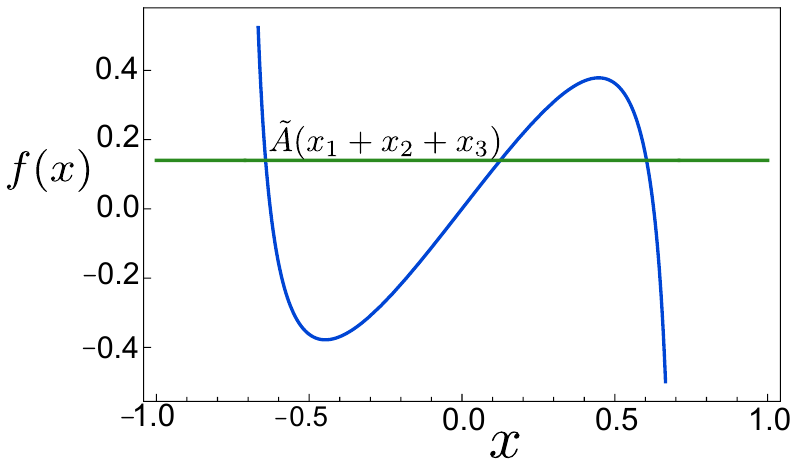}
        \label{fig:figure3}
    }
    \caption{The monotonicity of the function $f(x)$ for (a) $g<g_c^+$ and (b) $g>g_c^+$. The blue line represents the function of $f(x)$, the red line indicates the impossible solutions, and the green line indicates the possible solutions.}
    \label{fig:figures2}
\end{figure}

The ground state energy global minimum requires two first order derivatives Eq.~\eqref{eq:7_7} and Eq.~\eqref{eq:8} of  $\bar{E}_{GS}$ vanish, and we transform equation $\frac{\partial\bar{E}_{GS}}{\partial\theta_n}=0$ with the solution of $\frac{\bar{E}_{GS}}{{\partial\bar{\alpha}_n}}=0$ and $x_n$. After arrangement, we have 
\begin{equation} \label{eq:13}
\tilde{A}(x_1+x_2+x_3)=\frac{(g^2+J_2-J_1J_2)x}{1-J_1}-\frac{x}{\sqrt{1-\frac{4x^2}{g^2}}},
\end{equation}
where $\tilde{A}$=$\frac{J_2+J_1(g^2+J_2-2J_1J_2)
)}{(1+2J_1)(1-J_1)}$ is the coefficient of $\sum_{i=1}^3x_i$, and $x$ on the right side of the equation is without the subscript. It means $x$ to be any of $x_1, x_2$ or $x_3$ due to the periodic boundary condition. The right side of Eq.~\eqref{eq:13} can be regarded as a function $f$ of $x$ with the left side being a quantity $k$, and $x_{1,2,3}$ satisfying the above equation can be considered as the roots of $f(x)=k$. The function $f(x)$ is monotonic for $g<g_c^+$ and becomes non-monotonic for $g<g_c^+$, as shown in Fig.~\ref{fig:figures2}. Analyzing the intersection between $f(x)$ and $k=\tilde{A}(x_1+x_2+x_3)$, we can find out the possible solution for both the normal phase and the superradiant phase based on the monotonicity.


\section{Cauchy-Schwarz inequalities}\label{Appendix C}

For $(J_1$, $J_2)$ beneath the curve in the $J_1$-$J_2$ diagram in Fig.~\ref{fig:1} corresponding to the critical coupling strength $g_c^-$, we can find an analytic solution of mean values by investigating the ground state energy Eq.~\ref{eq:10} at the global minimum. We introduce two Cauchy-Schwarz inequalities,
\begin{equation} \label{eq:20}
\sum_{i=1}^3\sqrt{1-\frac{4x_i^2}{g^2}}\leq 3\sqrt{1-\frac{4}{3g^2}\sum_{i=1}^3x^2_i},
\end{equation}
\begin{equation} \label{eq:21}
x_1x_2+x_2x_3+x_3x_1\leq x_1^2+x_2^2+x_3^2,
\end{equation}
where the equality holds if and only if $x_1=x_2=x_3$ and $x_1^2=x_2^2=x_3^2$, respectively. With the interacting term's coefficient $\tilde{B}=(\frac{J_1}{1+J_1-2J_1^2}+\frac{J_2}{g^2})$ stays negative in Eq.~\ref{eq:10}, we can find the bottom limit of ground state energy 
\begin{equation} \label{eq:22}
\begin{split}
\bar{E}_{GS} &\geq\tilde{C}\sum_{i=1}^3x^2_i-\frac{3}{2}\sqrt{1-\frac{4}{3g^2}\sum_{i=1}^3x^2_i}+2\tilde{B}\sum_{i=1}^3x_i x_{i+1}
\\&\geq(\tilde{C}+2\tilde{B})\sum_{i=1}^3x^2_i-\frac{3}{2}\sqrt{1-\frac{4}{3g^2}\sum_{i=1}^3x^2_i},
\end{split}
\end{equation}
where the equality holds if and only if $x_1=x_2=x_3$.The condition that $\tilde{B}$ stays positive can be satisfied by any $(J_1,~J_2)$ above the curve in $J_1$-$J_2$ diagram in Fig.~\ref{fig:1} corresponding to the critical coupling strength $g_c^-$ in the superradiant regime. The global minimum of ground state energy can be obtained by minimizing $\bar{E}_{GS}$ with $\bar{\alpha}_1=\bar{\alpha}_2=\bar{\alpha}_3$ remaining the translational symmetry,
\begin{equation} \label{eq:23}
 \bar{\alpha}_{1,2,3}=\frac{1}{2} g\sqrt{\frac{1}{(1 + 2 J_1)^2} -\frac{1}{(g^2 - 2 (J_2 + 2 J_1 J_2))^2}}.
\end{equation}


\section{Dicke Trimer model with the atom hopping $J_2$ alone}

The Hamiltonian of the Dicke trimer model only with atom hopping $J_2$ is represented as
\begin{equation} \label{eq:132}
\tilde{H_2}=\sum_{i=1}^{3}\left[H^{\rm Dicke}_n+\frac{J_2}{N_a} \left(J^+_n J^-_{n+1}+J^-_{n} J^+_{n+1}\right)\right].
\end{equation}
Its ground state energy reads as
\begin{equation} \label{eq:132}
\begin{split}
 \bar{E}_{2} &=\sum^N_{n=1}(\bar{\alpha}_n^2+\frac{1}{2}\cos\theta_n+g\bar{\alpha}_n\sin\theta_n\cos\phi_n\\
&+
\frac{1}{2}J_2(\sin\theta_n\sin\theta_{n+1}\cos\phi_n\cos\phi_{n+1})
\end{split}
\end{equation}
after transformation and rescale. The ground state is determined by globally minimizing $\bar{E}_{2}$ with respect to $\bar{\alpha}_n$, namely,
\begin{equation} \label{eq:7}
\frac{\partial\bar{E}_{2}}{\partial\bar{\alpha}_n}=2\bar{\alpha}_n+g\sin\theta_n\cos\phi_n=0.
\end{equation}
Then, we obtain $\sin\theta_n=-2\alpha/g\cos\phi_n$ and $\cos\theta_n=-1/\sqrt{1-4\alpha_n^2/g^2}$. We use these solutions to integrate out the atom degrees of freedom and obtain the mean-field ground state energy $\bar{E}_{2}(\{\bar{\alpha}_{n}\})$, which reads as 
\begin{equation} \label{eq:77}
\bar{E}_{2} =\sum^N_{n=1}\left(-\bar{\alpha}_n^2-\frac{1}{2}\sqrt{1-4 \alpha_n^2/g^2}+\frac{2J_2}{g^2}\alpha_n \alpha_{n+1}\right).
\end{equation}
Studying this ground state energy $\bar{E}_{2}$ with the methods mentioned above, it turns out that the system undergoes frustrated superradiant phase transitions with hopping energy $J_2>0$ and normal superradiant phase transition with ${J_2}<0$, which has the same mechanism compared with the case that Dicke trimer interconnected only via photon hopping ${J_1}$~\cite{PhysRevLettFSPT}.


\section{Linear Relation between $\alpha$ and $x$}\label{AppendixE}

Based on the linear relation $x_n=\bar{\alpha}_n+J_1(\bar{\alpha}_{n-1}+\bar{\alpha}_{n+1})$, we have
\begin{equation} \label{eq:324}
  \left( \begin{array}{c}
        x_{n-1} \\
        x_{n} \\
        x_{n+1}
        \end{array}\right)=
  \left( \begin{array}{ccc}
        1 & J_1 &J_1 \\
        J_1 & 1 &J_1 \\
        J_1 & J_1 & 1
        \end{array}\right)
          \left( \begin{array}{c}
        \bar{\alpha}_{n-1} \\
        \bar{\alpha}_{n} \\
         \bar{\alpha}_{n+1}
        \end{array}\right).
\end{equation}
This expression can be written in a concise way as 
\begin{equation} \label{eq:324}
 X=S A
\end{equation}
with $X=(x_{n-1},x_{n},x_{n+1})^T$ and $A=(\alpha_{n-1},\alpha_{n},\alpha_{n+1})^T$. Since $S$ is a symmetric matrix, the off-diagonal elements satisfy $S_{12}=S_{21}, S_{13}=S_{31}$, and $S_{23}=S_{32}$. This means that the relations between $x_{n-1}$, $x_{n}$, $x_{n+1}$, and $\bar{\alpha}_{n-1}$, $\bar{\alpha}_{n}$, $\bar{\alpha}_{n+1}$ are not asymmetric, repectively. Due to the symmetry of $S$, the inverse matrix $S^{-1}$ is also symmetric and we have
\begin{equation} \label{eq:324}
 A=S^{-1} X,
\end{equation}
which means $\bar{\alpha}_{n-1}$, $\bar{\alpha}_{n}$, $\bar{\alpha}_{n+1}$ are symmetric in form to the equations for $x_{n-1}$, $x_{n}$, $x_{n+1}$ establishing a identical relation.

\bibliography{references}

\end{document}